\documentclass[12pt,preprint]{aastex}
\slugcomment{2012.7 v2}

\shorttitle{Spin down of SGR 0418+5729} \shortauthors{Tong \& Xu}

\begin{document}

\title{SGR 0418+5729: a small inclination angle resulting in a not so low dipole magnetic field?}

\author{H. Tong\altaffilmark{1}, R. X. Xu\altaffilmark{2}}

\altaffiltext{1}{Xinjiang Astronomical Observatory, Chinese Academy
of Sciences}
\altaffiltext{2}{School of Physics and State Key
Laboratory of Nuclear Physics and Technology, Peking University}

\begin{abstract}
The spin down behaviors of SGR 0418+5729 are investigated.
The pulsar spin down
model of Contopoulos \& Spitkovsky (2006) is applied to SGR 0418+5729. It is shown that SGR 0418+5729 lies below the
pulsar death line and its rotation-powered magnetospheric activities may therefore have stopped. 
The compact star is now
spun down by the magnetic dipole moment perpendicular to its rotation axis. Our calculations show that 
under these assumption there is the possibility of SGR 0418+5729 having a strong dipole magnetic field, if there is
a small magnetic inclination angle.
Its dipole magnetic field may be much higher than the characteristic magnetic field.
Therefore, SGR 0418+5729 may  be a normal magnetar instead of a low magnetic field magnetar.

\end{abstract}


\section{Introduction}

Since the discovery of pulsars, different manifestations of pulsar-like objects have been observed. Among them,
anomalous X-ray pulsars (AXPs) and soft gamma-ray repeaters (SGRs) are two kinds of enigmatic
pulsar-like objects. AXPs and SGRs are magnetar candidates, i.e. magnetism-powered neutron stars.
During their studies, the magnetic dipole braking assumption are often employed. A dipole magnetic field
 larger than the quantum critical value
($B_{\rm QED}\equiv 4.4\times 10^{13} \,\rm G$)
is often taken as confirmation of star's magnetar nature (Kouveliotou et al. 1998). The traditional
picture of magnetars is: they are neutron stars with both strong dipole field and strong multipole field
(Thompson et al. 2002; Mereghetti 2008; Tong \& Xu 2011).

This traditional picture of magnetars is challenged by the discovery of a so-called ``low magnetic field'' magnetar
SGR 0418+5729 (Rea et al. 2010). According to Rea et al. (2010), SGR 0418+5729 has a rotation period
$P=9.08 \,\rm s$, and a period derivative $\dot{P}<6.0\times 10^{-15}$. Therefore, its dipole magnetic field
is $B_{\rm c}<7.5\times 10^{12} \, \rm G$, assuming magnetic dipole braking. The dipole magnetic field of SGR 0418+5729
is much lower than the quantum critical value. Therefore, it challenges the traditional picture of magnetars
(Rea et al. 2010). If the characteristic magnetic field is star's true dipole magnetic field, then there
may be significant magnetic field decay during the life time of SGR 0418+5729 (Turolla et al. 2011).
Furthermore, it means that radio pulsars can also show magnetar-like bursts (Perna \& Pons 2011).

However, the dipole magnetic field of SGR 0418+5729 is obtained by assuming magnetic dipole braking.
For pulsars near the death line, their dipole magnetic field can be much higher than
the characteristic magnetic field, according to the pulsar spin down model of Contopoulos \&
Spitkovsky (2006, hereafter CS2006). In this paper,
we apply the pulsar spin down model of CS2006 to SGR 0418+5729.
Our calculations show that under these assumptions, the dipole magnetic field of SGR 0418+5729 may still be very strong, 
much higher than $10^{13} \,\rm G$.

Model calculations are given in Section 2. Discussions are presented in Section 3.

\section{Modeling the spin down of SGR 0418+5729}

\subsection{Description of pulsar spin down models}

Both normal pulsars and magnetars are often assumed to be braked down via magnetic dipole radiation.
Their characteristic magnetic field and characteristic age are calculated in this way.
The ``magnetic dipole braking'' is calculated for an orthogonal rotator in vacuum.
The magnetic inclination angle are taken to be $90^{\circ}$ in calculating the characteristic magnetic field
(Lyne \& Graham-Smith 2012, eq.(5.17) there).

The general case should be an oblique rotator surrounded
by plasmas. In the vicinity of the star, acceleration gaps are formed (Li et al. 2012; Kalapotharakos et al. 2012).
For an oblique rotator, the magnetic moment can be decomposed into two components: one perpendicular
to the rotation axis and the other parallel to the rotation axis. Therefore, the electromagnetic spin
down torque will be a combination of magnetic dipole radiation and particle outflow (Xu \& Qiao 2001; CS2006).
The analytical treatment of CS2006 is confirmed by numerical simulation of pulsar magnetospheres
(Spitkovsky 2006). For pulsars above the death line, they are quantitatively similar. However, Spitkovsky (2006)
is for the force-free magnetosphere. It does not include the existence of acceleration gaps. 
For pulsars near/below the death line, the particle outflow component will cease to operate. This point is considered in CS2006, 
while it can not be modeled in Spitkovsky (2006). Recent numerical simulations taken into consideration the effect
of acceleration gaps also find similar results to that of CS2006 (eq.(13) in Li et al. 2012 and corresponding discussions).
Therefore, we employ the analytical treatment of CS2006 and apply it to SGR 0418+5729, for the sake of simplicity.

According to the CS2006 (eq.(8) there), the electromagnetic spin down torque is
\begin{eqnarray}\label{Lem}
L&=& L_{\rm orth} \sin^2\theta + L_{\rm align} \cos^2\theta \nonumber\\
 &=& \frac{B_{\ast}^2 \Omega^2 r_{\ast}^6}{4 c r_{\rm c}^2}
 [ \sin^2\theta + (1- \frac{\Omega_{\rm death}}{\Omega}) \cos^2\theta ].
\end{eqnarray}
Here $L_{\rm orth}$ is the electromagnetic torque in the orthogonal case,
$L_{\rm align}$ the electromagnetic torque in the aligned case,
$\theta$ the angle between magnetic moment and rotation axis (the magnetic inclination angle),
$B_{\ast}$ the surface dipole magnetic field (at the magnetic pole),
$\Omega$ the stellar angular rotation frequency,
$r_{\ast}$ the stellar radius,
$c$ the speed of light,
$r_{\rm c}$ the radial extension of the closed field line regions,
$\Omega_{\rm death}$ the pulsar death period. The spin down torque of an oblique rotator
is the combination of the orthogonal torque and the aligned torque. For pulsars near the death line $\Omega \sim \Omega_{\rm death}$,
the align component will be stopped. The star is slowed down mainly by magnetic dipole radiation with an
effective magnetic field $B_{\ast}\sin\theta$. Therefore, for pulsars near the death line, their dipole magnetic field will be much higher than
their characteristic magnetic field if the star has a small inclination angle.
This may be the case of SGR 0418+5729.

\subsection{Spin down of SGR 0418+5729}

\begin{figure}[!ht]
 \centering
\includegraphics{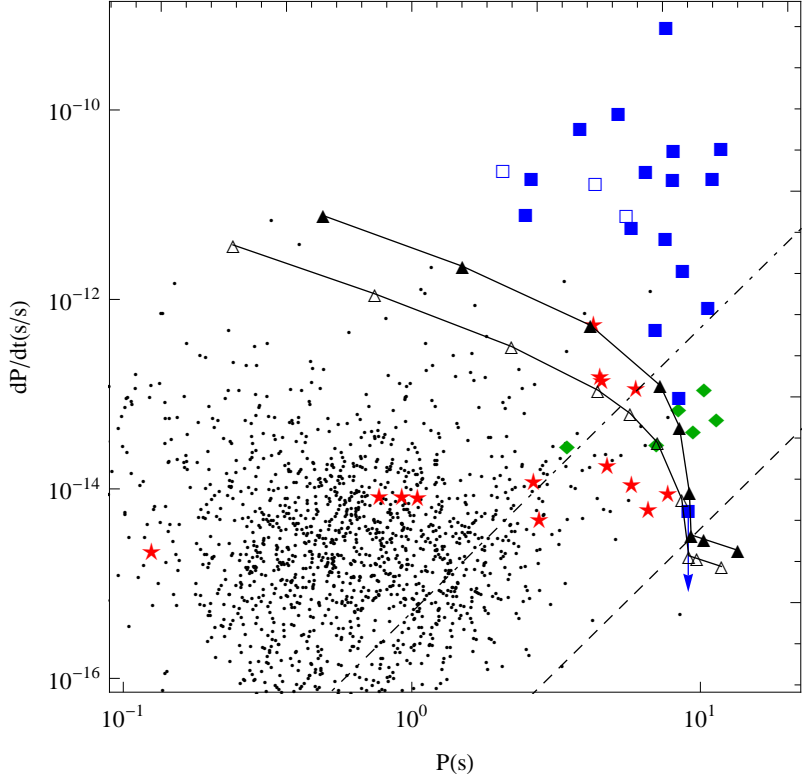}
\caption{Distribution of magnetars on the $P-\dot{P}$ diagram.
Squares are for magnetars, while empty squares are radio emitting magnetars,
the down arrow marks the position of SGR 0418+5729
(McGill online catalog: http://www.physics.mcgill.ca/$\sim$pulsar/magnetar/main.html).
Diamonds are for X-ray dim isolated neutron stars (Kaplan \& van Kerkwijk 2011).
Stars are for rotating radio transients, dots are for normal pulsars (ATNF:
http://www.atnf.csiro.au/research/pulsar/psrcat/).
The dot-dashed line is the pulsar death line for an orthogonal rotator (assuming $V_{\rm gap}=10^{13} \,\rm V$). 
The dashed line is the corresponding death line for pulsars with inclination angle $5^{\circ}$.
The triangles are model calculations of spin down evolution of SGR 0418+5729. See text for details. }
\label{fig0418}
\end{figure}

The radial extension of closed field line regions can be taken as the light cylinder
radius $r_{\rm lc}=c/\Omega$. This corresponds to power index $\alpha=0$ in eq.(11) in CS2006. The spin down behavior
can be obtained by setting $\alpha=0$ in the corresponding expressions in CS2006\footnote{The power index
of $\left( \frac{P_0}{1 \,\rm s} \right)$ in eq.(13) in CS2006 should be $-\alpha/(2-\alpha)$.}. The period derivative is
(eq.(12) in CS2006)
\begin{eqnarray}\label{Pdot}
\dot{P} &=& 3.3\times 10^{-16} \left( \frac{P}{1\,\rm s} \right)^{-1} \left( \frac{B_{\ast}}{10^{12} \, \rm G} \right)^{2}
[\sin^2\theta + \left(1- \frac{P}{P_{\rm death}}\right) \cos^2\theta], \quad  \mbox{for $P\le P_{\rm death}$}  \nonumber \\
&=& 3.3\times 10^{-16} \left( \frac{P}{1\,\rm s} \right)^{-1} \left( \frac{B_{\ast}}{10^{12} \, \rm G} \right)^{2}
\sin^2\theta, \quad  \mbox{for $P>P_{\rm death}$}.
\end{eqnarray}
The death period $P_{\rm death}$ is the maximum rotation period of a pulsar in order to maintain a constant gap potential. For $P>P_{\rm death}$,
the pair production and the pulsar rotation-powered magnetospheric activities are stopped.
The pulsar death period is (eq.(13) in CS2006)
\begin{equation}\label{Pdeath}
P_{\rm death} = 2.8 \left( \frac{B_{\ast}}{10^{12} \,\rm G} \right)^{1/2} \left( \frac{V_{\rm gap}}{10^{12} \,\rm V} \right)^{-1/2} \,\rm s,
\end{equation}
where $V_{\rm gap}$ is the potential drop in the acceleration gap. When $P=P_{\rm death}$, the corresponding
period derivative is (i.e. the pulsar death line, eq.(14) in CS2006)
\begin{equation}\label{deathline}
\dot{P}_{\rm death} = 5\times 10^{-18} \left( \frac{P_{\rm death}}{1 \,\rm s} \right)^{3} \left( \frac{V_{\rm gap}}{10^{12} \,\rm V} \right)^{2}
\sin^2\theta.
\end{equation}
The distribution of magnetars on the $P-\dot{P}$ diagram is shown in figure \ref{fig0418}. We also plot the death line for an orthogonal rotator
in figure \ref{fig0418}.

From figure \ref{fig0418} we see that SGR 0418+5729 lies far below the death line. Therefore, its
roation-powered magnetospheric activities may have already stopped. Its X-ray emissions are
magnetism-powered. If we assume that the braking mechanism of SGR 0418+5729
is similar to that of rotation-powered pulsars, then there are two possibilities
concerning the spin down of SGR 0418+5729:
\begin{enumerate}
  \item It has a large inclination angle (e.g. $\theta>45^{\circ}$). The characteristic magnetic field
  is a good estimate of its true dipole magnetic field. Then it will indeed be a low magnetic field magnetar, and
  its age is relatively large ($>2.4\times 10^7 \,\rm yr$, Rea et al. 2010). During its life time, its
  dipole magnetic field have decayed significantly (Turolla et al. 2011).

  \item It has a small inclination angle (e.g. $\theta \simeq 5^{\circ}$). Its dipole magnetic field will be
  much higher than the characteristic magnetic field. Its true age will also be smaller than the characteristic age.
  In this case, SGR 0418+5729 will be a normal magnetar.
  It has a small period derivative because its magnetic inclination is small.
\end{enumerate}

If SGR 0418+5729 has a small inclination angle, we can estimate its dipole magnetic field and inclination angle in the following ways.
Considering its position on the $P-\dot{P}$ diagram, its period will be close to its death period $P\approx P_{\rm death}$.
From eq.(\ref{Pdeath}), the dipole magnetic field is $B_{\ast} \approx 10^{14} (V_{\rm gap}/10^{13} \,\rm V) \,\rm G$.
Typically, $V_{\rm gap}=10^{13} \,\rm V$ is used in CS2006 for normal pulsars. The period derivative of SGR 0418+5729
will be larger than that at $P=P_{\rm death}$. From eq.(\ref{Pdot}), the upper limit on inclination angle is
$\theta< 7^{\circ} (V_{\rm gap}/10^{13} \,\rm V)^{-1}$.

From eq.(\ref{Pdot}), given a set of $(B_{\ast}, \theta, V_{\rm gap})$,
we can calculate the spin down evolution of SGR 0418+5729.
For $(B_{\ast}=1.1\times 10^{14} \,\rm G, \theta=5^{\circ}, V_{\rm gap}=10^{13} \,\rm V)$, the spin down evolution of SGR 0418+5729
is shown in figure \ref{fig0418} as filled triangles. The period and period derivative are shown for pulsar age:
$(10^3 \,\rm yr, 10^4 \,\rm yr, 10^5 \,\rm yr, 5\times 10^5 \,\rm yr, 10^6 \,\rm yr, 2\times 10^6 \,\rm yr, 2.7\times 10^6 \,\rm yr, 10^7 \,\rm yr, 5\times 10^7 \,\rm yr)$, respectively. The data point of $t=2.7\times 10^6 \,\rm yr$
corresponds to pulsar death point. The empty triangles in figure \ref{fig0418} is for $(B_{\ast}=5.3\times 10^{13} \,\rm G, \theta=8^{\circ}, V_{\rm gap}=5\times 10^{12} \,\rm V)$. The period and period derivative data points are for pulsar age:
$(10^3 \,\rm yr, 10^4 \,\rm yr, 10^5 \,\rm yr, 10^6 \,\rm yr, 2\times 10^6 \,\rm yr, 5\times 10^6 \,\rm yr, 8.7 \times 10^6 \,\rm yr,
10^7 \,\rm yr, 5\times 10^7 \,\rm yr)$, respectively. The data point of $t=8.7 \times 10^6 \,\rm yr$
is the pulsar death point. When the pulsar passes the death point, the parallel component of spin down torque is stopped.
The star will be spun down by pure magnetic dipole radiation, under the influence of an effective magnetic field $B_{\ast}\sin\theta$
(CS2006).
From the model calculations, we see that there are still parameter space that
SGR 0418+5729 has a much higher dipole magnetic field. SGR 0418+5729 may still be a normal magnetar instead of a low magnetic field magnetar.

If the dipole magnetic field of SGR 0418+5729 is really much higher, e.g. $\approx 10^{14} \,\rm G$, it will be more burst active (Perna \& Pons 2011).
According to CS2006, for stars near the death line with small inclination angles, they will have a very large braking index.
However, this is only for sources before they pass the death point. It is also possible that SGR 0418+5729 has already passed
the death point and it is now spun down by pure magnetic dipole radiation.
The braking index in this case will be $n=3$. At present we only know the period derivative upper limit of SGR 0418+5729.
Future exact period derivative measurement and even braking index measurement of this source can tell us whether it has
passed the death point or not.

\section{Discussions}

In this paper, we employ the pulsar spin down model
of CS2006. In CS2006, the aligned torque is proportional to $1-\Omega_{\rm death}/\Omega=1-V_{\rm gap}/V_{\ast}$, where $V_{\ast}$ is the
polar cap potential drop (CS2006). Similar dependence is also found by up-to-date numerical simulations of pulsar magnetospheres 
(eq.(13) in Li et al. 2012). 
Compared with the results of numerical simulations, the CS2006 model involves an additional angular dependence factor $\cos^2\theta$.
For the small inclination angle case we considered here, $\cos^2\theta \approx 1$. Therefore, our calculations are insensitive to this
angular dependence factor. Future more detailed numerical simulations may improve some of the numerical factors (e.g. eq.(13) in Li et al. 2012). 
However, the physical picture as outlined in CS2006 may always exist: (1) The pulsar spin down torque is the combination of an orthogonal 
component (magnetic dipole radiation) and parallel component (particle outflow). (2) Near the death-line, the parallel component will 
be ceased, but only the orthogonal component survives. In conclusion, the model of CS2006 is consistent with up-to-date numerical simulations (Li et al. 2012).
Therefore, we prefer to employ the analytical model of CS2006. 

Radio observations of magnetars have shown that the three radio emitting magnetars may all have nearly aligned geometry
(Camilo et al. 2007, 2008; Levin et al. 2012). Therefore, a small inclination angle for SGR 018+5729 (e.g. $\approx 5^{\circ}-10^{\circ}$)
is not impossible. A small inclination angle can also not be ruled out by present X-ray observations (see Esposito et al. 2010
for discussions and references therein).
If SGR 0418+5729 has a small inclination angle, then considering its position in the $P-\dot{P}$ diagram, 
the aligned component of its spin down torque may have already stopped. 
It is now spun down mainly by magnetic dipole radiation under the effective magnetic field $B_{\ast} \sin\theta$. 
Therefore, the characteristic magnetic field may significantly under estimate its true dipole magnetic field. 
This result is insensitive to the detailed expression of pulsar spin down torque. 
If SGR 0148+5729 has a small inclination angle, its dipole magnetic field will be much higher than $10^{13}\,\rm G$. Its true age 
will be smaller than the characteristic age. Therefore, it will be more burst active (Perna \& Pons 2011). 
If it still has not passed the death point, then it will have a very large braking index (CS2006). 
These predictions can be tested by future observations.

The radial extension of closed field line regions is taken as the light cylinder radius. This is the case for most normal pulsars
according to CS2006. By setting $r_{\rm c} =r_{\rm lc}$, the corresponding spin down torque is also consistent with results of numerical
simulations (Spitkovsky 2006; Li et al. 2012). Except when the braking mechanism of magnetars differ qualitatively from that
of normal pulsars (e.g. strong wind braking), the light cylinder radius will be the natural length for $r_{\rm c}$. 
In the case of wind braking of magnears (Tong et al. 2012), $r_{\rm c}$ will be smaller than the light cylinder radius. 
The dipole magnetic field will correspondingly
be smaller. We do not consider this possibility here. The above discussions and calculations are done under the assumption that
the braking mechanism of SGR 0418+5729 is similar to that of rotation-powered pulsars.

From figure \ref{fig0418}, the second low magnetic SGR Swift J1822.3-1606 (Rea et al. 2012) and several X-ray dim isolated neutron stars
lie also near the death line. It is possible that their dipole magnetic field is also higher than the characteristic magnetic field.
However this effect will not be so significant as in the case of SGR 0418+5729, which lies far below the death line. As has already
been suggested in CS2006, we also hope future population synthesis of magnetars and X-ray dim isolated neutron stars to take
this geometrical effect into consideration.

The persistent X-ray luminosity of SGR 0418+5729 is similar to that of AXP XTE J1810--197 (e.g. figure 1 in Tong et al. 2012
and references therein).
They are both lower than the rest of magnetars. However, the period derivative of SGR 0418+5729 is three orders of magnitude
smaller than that of XTE J1810--197. It may be that the dipole magnetic field of SGR 0418+5729 is much smaller than XTE J1810--197. 
In this case, XTE J1810--197 is a normal magnetar, while SGR 0418+5729 will be a low magnetic field magnetar. This is the commonly assumed case. 
Another possibility is: SGR 0418+5729 has a small inclination angle, while its dipole magnetic field is higher than the quantum
critical value. 
XTE J1810--197 lies above the death line (see figure \ref{fig0418}).
Therefore, irrespective of its inclination angle, its spin down torque will always be very efficient.
Meanwhile, for SGR 0418+5729 which lies far below the
death line, it will mainly be spun down under the effective magnetic field $B_{\ast} \sin\theta$. For small inclination
angle case, the consequent period derivative will be very small.

In conclusion, considering the detailed modeling of pulsar spin down torque of CS2006, 
it is possible that SGR 0418+5729 has a strong dipole magnetic field,
if there is a small inclination angle. It may be a normal magnetar instead of a low magnetic field magnetar.
Future observations may help us to distinguish between these two possibilities.
Making clear this problem will also test one of the basic assumptions in magnetar researches, i.e.
the magnetic dipole braking assumption.

\section*{Acknowledgments}

This work is supported by the National Natural Science Foundation
of China (Grant Nos. 10935001, 10973002, 11103021), the National Basic Research Program of China
(Grant No. 2009CB824800), and the John Templeton Foundation.

\end{document}